\documentclass[prb,aps,twocolumn,showpacs,amsmath,amssymb,preprintnumbers,floats,floatfix]{revtex4}

\usepackage{graphicx}
\usepackage{dcolumn}
\usepackage{amssymb}
\usepackage{bm}
\usepackage[T1]{fontenc}
\usepackage{t1enc}
\usepackage{textcomp}
\usepackage{indentfirst}

\begin{document}

\title{Measurement of the separation dependence of resonant energy transfer between CdSe/ZnS core/shell nanocrystallite quantum dots}

\author{Farbod Shafiei}
\address{Department of Physics, Indiana University-Purdue University Indianapolis, Indianapolis IN 46202}

\author{Sannah P. Ziama}
\address{Department of Physics, Indiana University-Purdue University Indianapolis, Indianapolis IN 46202}

\author{Eric D. Curtis}
\address{Department of Physics, Indiana University-Purdue University Indianapolis, Indianapolis IN 46202}

\author{Ricardo S. Decca}
\email{rdecca@iupui.edu}
\address{Department of Physics, Indiana University-Purdue University Indianapolis, Indianapolis IN 46202}

\date{\today}

\begin{abstract}
The separation dependence of the interaction between two resonant
groups of CdSe/ZnS nanocrystallite quantum dots is studied at room
temperature. A near-field scanning optical microscope is used to
bring a group of mono-disperse $\sim $6.5 nm diameter
nanocrystallite quantum dots which are attached to the microscope
probe, into close proximity of $\sim$8.5 nm diameter group of
nanocrystallite quantum dots which are deposited on a solid
immersion lens. Information extracted from photoluminescence,
photoluminescence excitation and absorption curves as well as
numerical calculations of the energy levels, show that the third excited
excitonic energy level of the large quantum dots nearly matches
the ground excitonic energy level for the small quantum dots.
Quenching of the small quantum dots photoluminescence signal has
been observed as they approach the large quantum dots. On average,
the separation between microscope probe and solid immersion lens changed in the
15-50 nm range. The transition probability between these two groups
of quantum dots is calculated to be
$(2.60\times10^{-47}m^{6})/R^{6}$, within the
$(0.70\times10^{-47}m^{6})/R^{6}-(11.0\times10^{-47}m^{6})/R^{6}$
experimentally obtained range of transition probabilities. The
F\"{o}rster radius, as a signature of energy transfer efficiency,
is experimentally found to be in the 14-22 nm range.
\end{abstract}


\maketitle

\section{\label{sec:level1}Introduction}
Advanced semiconductor technology starting in late 1980s allowed for
the fabrication of nanocrystallite quantum dots (NQDs), consisting
of a few hundred to many thousand atoms\cite{Allivisatos1996} of
semiconductor materials producing a potential well for electrons
and holes. NQDs are fabricated such that their diameters are
smaller than the bulk Bohr exciton diameter, thus the electronic
structure is dominated by quantum confinement effects in all three
dimensions \cite{Efros1982,Brus1984,Reimann2002} and is suited for
the study of zerodimensional structures.
\cite{Allivisatos1996,Brus1991,Empedocles1996} Colloidal NQDs,
which are synthesized by relatively inexpensive wet chemistry
methods, have high control in engineering the energy levels. This
results in NQDs with strong size dependent optical and electrical
properties. \cite{Allivisatos1996} In particular, CdSe NQDs can be
synthesized with a tunable size of 15-100~{\AA} in a narrow
distribution ($<$~5$\%$rms dispersion). \cite{Murray1993}

Emission properties of NQDs are often measured via
photoluminescence (PL) experiments. In PL, excitonic states in the
semiconductor material are induced by photon absorption, and the
optical emission as these excitons recombine analyzed. In F\"{o}rster
resonant energy transfer (FRET) an excited donor can transfer its
energy directly (nonradiatively) to an acceptor \textit{via}
dipole-dipole interaction. The phenomenon of resonant energy
transfer was observed by J. B. Perrin \cite{Wu1994,Masters2008} at the beginning of the 20th
century, but it was F\"{o}rster in the late 1940s
\cite{Forster1948} who proposed a theory describing long range
molecular interaction by resonance energy transfer. Due to its strong separation
dependence, FRET has been used as a molecular ruler to determine
inter- and intra-molecular distances.\cite{Stryer1999} Since FRET
represents a transfer of energy, it can be detected by measuring
the quenching of donor emission or the enhancement of
acceptor emission. This relationship of the transfer
rate as a function of donor-acceptor separation was first
demonstrated with peptides in 1963.\cite{Edelhoch1963}

Controlling the distance between the NQDs in real time has been a
challenge faced by most groups studying the dynamic process of
energy transfer between
NQDs.\cite{Kagan1995,Kagan1996,Crooker2002,Nomura2007,Chen2008,Lu2008,Kim2008,Kim2009,Wang2010,Tai2010}
We use a near-field scanning optical microscope (NSOM) to control
the distance between two groups of NQDs in real time. In this experiment a
group of core/shell CdSe/ZnS NQDs with external diameter of
$\sim$~6.5 nm, attached to the apex of the NSOM probe, are brought
into close proximity to a second group of CdSe/ZnS NQDs with
external diameter of $\sim$~8.5 nm, which are deposited on the
flat part of a solid immersion lens (SIL). Both groups of NQDs are
excited and the PL signal of the small NQDs is monitored to
observe any changes. Using an Al-coated NSOM probe and diluting
the NQDs helps to reduce the number of NQDs excited on the NSOM
probe and SIL.

\begin{figure}[h]
\begin{center}
\includegraphics[scale=0.35]{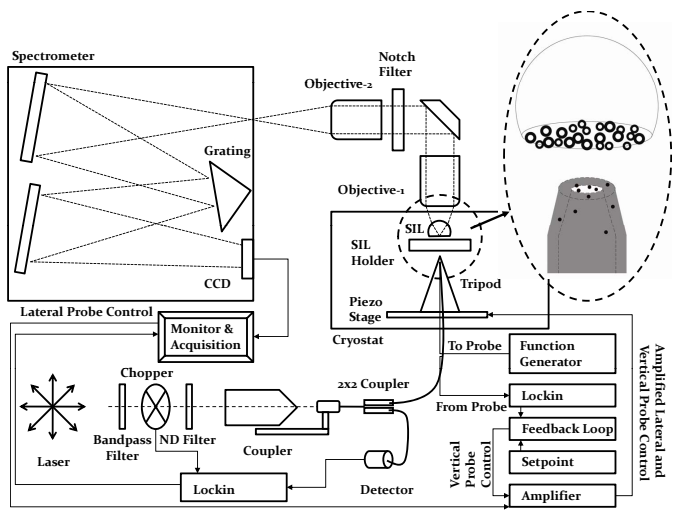}
\caption{Schematic of experimental setup including NSOM. Inset:
Schematic of NSOM probe and SIL (not to scale) with small NQDs on
the NSOM probe and large ones on the SIL. Distance between small and large
NQDs changes by moving the NSOM probe toward and away from the SIL. PL signals of
both groups of NQDs are collected on the CCD camera in the spectrometer.\label{fig:system}}
\end{center}
\end{figure}

\section{\label{sec:level1}Experiment}
\subsection{\label{sec:level2}Experimental setup}

An aperture NSOM has been designed and built to be used as a probe
to excite a small number of core/shell NQDs (\textrm{CdSe/ZnS}).
The main reason behind using a NSOM system was to first illuminate
an area much smaller than what can be achieved in far field
microscopy by overcoming the diffraction limit by the size of
probe.
\cite{Pohl1984,Harootunian1986,Durig1986,Betzig1987,Betzig1991,Betzig1992}
Second, the NSOM system can be used to move one group of NQDs with
respect to another, making the separation between them arbitrarily small by feedback schemes.
\cite{Pohl1984,Harootunian1986,Durig1986,Betzig1987,Betzig1991,Betzig1992,Karrai1995}
Small NQDs are attached to the apex of a NSOM
probe by dipping the probe into the colloidal suspension of the
NQDs, while large NQDs are diluted and deposited on the SIL by
drop cast. All NQDs are covered by octadecylamine (ODA) ligands.
The inset to Fig.~\ref{fig:system} schematically shows
the probe's Al-coating precluding the excitation of the NQDs
outside its apex.

As it is shown in Fig.~\ref{fig:system}, the probe's vertical
motion is controlled by a feedback loop system while its lateral
motion is computer controlled. The amplified signal from the
feedback loop system and the computer are applied to a 3-axis
piezo stage. The NSOM probe is assembled on a tripod which sits on
the piezo stage. The probe approaches the flat side of a SIL
through a hole on the SIL holder. While all the results reported
in this paper were obtained at room temperature, the SIL holder
also serves the purpose of cooling down the SIL as it is in
thermal contact with the cryostat's cold finger. An argon laser
(488nm line) has been used for optical excitation. To improve on the
signal-to-noise ratio of the monitored signal, the laser light is
chopped before being coupled into a single mode optical fiber. The
use of the $2\times2$ fiber splitter, a Si photodiode and standard
lock-in detection allows for continuous monitoring of the laser
intensity coupled to the NSOM probe. After excitation of small and
large dots on the NSOM probe and the SIL, photons released by
these two groups of dots are collected through the SIL and two
other objectives. They are then dispersed by a grating
spectrometer and their energy is recorded on a charged-coupled
device (CCD).

As the NSOM probe is brought close to the SIL, a portion of the
small NQDs' energy is expected to be transferred to the large NQDs
and the PL signal of small NQDs' should show quenching.

\begin{figure}[h]
\begin{center}
\includegraphics[scale=0.78]{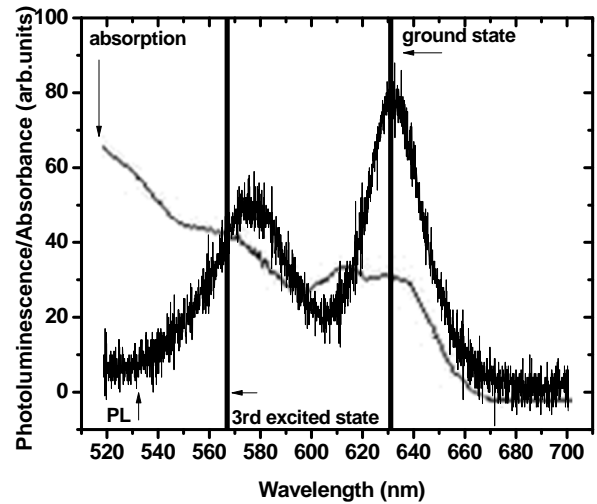}
\caption{Comparison of calculated energy levels, the absorbance spectrum of large NQDs and PL of resonant NQDs.
Vertical lines at 567 nm and 631 nm show numerically
calculated third excited and ground state energy levels of large
CdSe/ZnS NQDs with $\sim$~8.5 nm diameter. The fine line shows the absorbance
spectrum of large CdSe/ZnS NQDs provided by N.N. Labs LLC. This absorbance
spectrum shows that the large NQDs with $~630$ nm ground state energy has
an excited energy level also at $~570$ nm. PL signals from small NQDs
on the NSOM probe and large NQDs on the SIL excited by an argon laser
(488nm) are shown in a thick line. The calculated third excited energy level and absorption
spectrum of large NQDs at $~570$ nm matches the PL signal of the small NQDs.\label{fig:PL+Abs+Levels}}
\end{center}
\end{figure}

\subsection{\label{sec:level2}Measurement of the distance between small and large NQDs}

In the NSOM system, the amplitude of vibration of the NSOM coated
fiber probe, glued to a vibrating tuning fork driven at resonance,
has been used as an input for the feedback loop circuit.\cite{Karrai1995} This
circuit controls the distance between the NSOM probe and SIL.
The vibrational amplitude of the NSOM probe decreases as it is driven towards the SIL.
\cite{Karrai1995,Weiner1992,Toledo1992,Grober1994} This amplitude
damping has been used to measure the probe-SIL separation. To be
able to measure this distance, the probe was engaged in close
proximity of the SIL and then moved towards the SIL by decreasing
the setpoint in the feedback loop system. As the probe approaches
the SIL and its amplitude decreases, the system reaches the point
that its amplitude becomes unstable and the probe would break if
moved any further. This point is assumed to be the contact point
between the NSOM probe and the SIL. The separation is then
increased by pulling back the probe by increasing the setpoint.
Since the voltage applied to move the probe away from contact, as
well as the displacement calibration of the piezo stage as a
function of voltage, are known, the separation between the NSOM
probe and SIL can be obtained. Hence, in other experiments the voltage-separation
calibration curve has been used to identify the distance between
the small and large NQDs that are on the NSOM probe and SIL respectively.

\begin{figure}
\begin{center}
\includegraphics[scale=0.35]{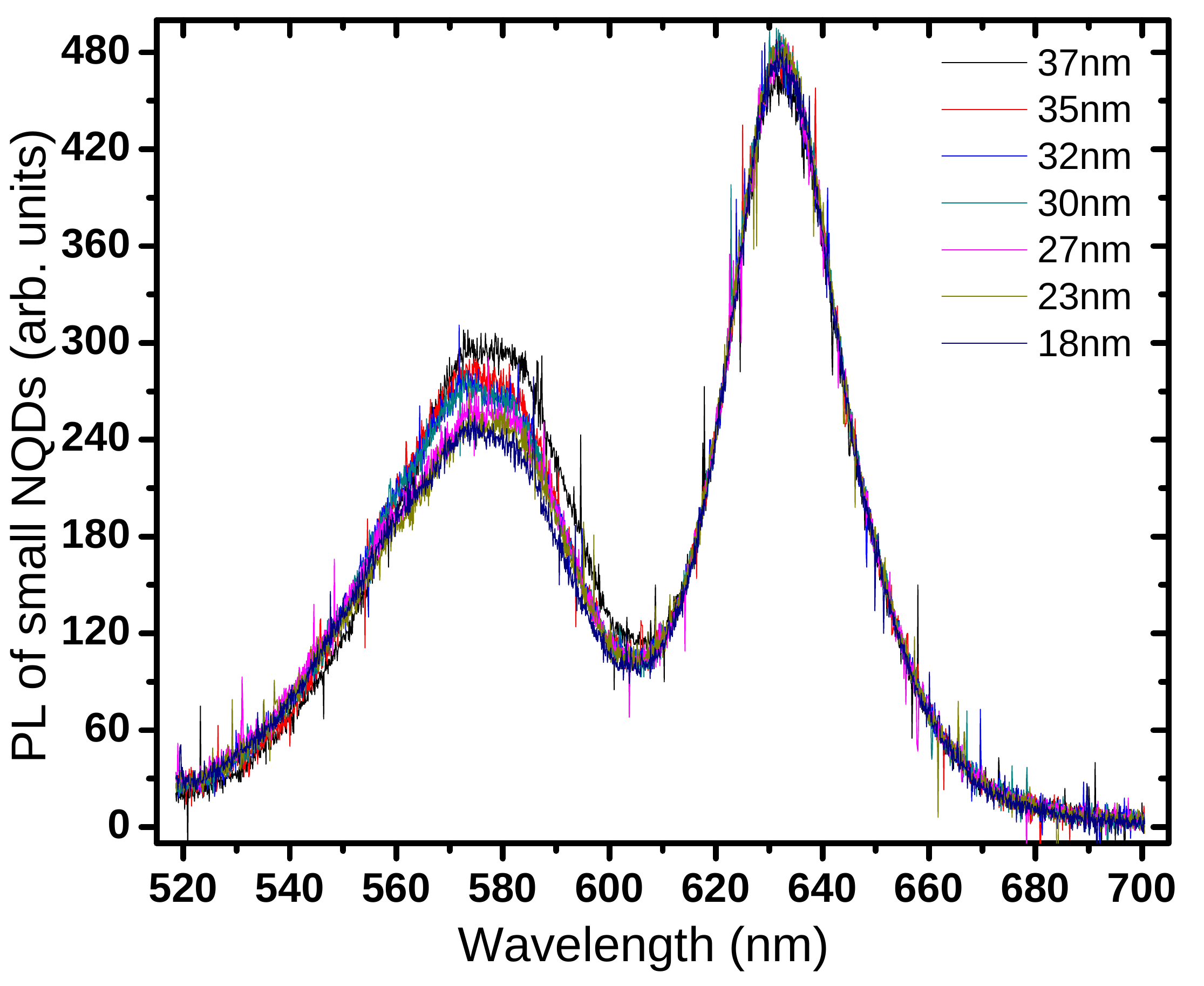}
\caption{(Color online) PL signal of small and large NQDs as a function of separation
between the NSOM probe and the SIL. The PL signal of small NQDs
decreases as the separation between the two groups of NQDs
decreases. The PL signal of small NQDs at left side from top to bottom corresponds to
37, 35, 32, 30, 27, 23 and 18 nm separation between NSOM probe and SIL.
Enhancement of large NQDs PL signal was not observed as
a consequence of the deposition process of the large NQDs on the
SIL, see Fig.\ref{fig:Image+Both+QDs}. \label{fig:PL}}
\end{center}
\end{figure}

\subsection{\label{sec:level2}Resonant CdSe$/$ZnS NQDs}

Specific sizes of small and large NQDs are selected to have the
excitonic ground state of the small CdSe/ZnS NQDs coincide with
one of the excited states of the large CdSe/ZnS NQDs. This energy
selection is accomplished by a numerical calculation of the energy
levels, and verified by PL, photoluminescence excitation (PLE),
and absorption experiments.
PL measurements were used to study the energy structure of the
NQDs by using the photon excitation and relaxation. It is shown in
the theory section that when the PL signal of the large NQDs with
$\sim$~8.5 nm diameter is observed at $\sim$~630nm, the calculated
corresponding PL signal for resonant set of small NQDs should be
at $\sim$~570nm, which corresponds to NQDs with $\sim$~6.5 nm
diameter. Vertical lines in Fig. \ref{fig:PL+Abs+Levels} show the
calculated third excited and ground state energy levels of large
NQDs at 567nm and 631nm.  To find a resonant pair of NQDs, PL
signals of various NQDs have been studied. As it is shown in
Fig. \ref{fig:PL+Abs+Levels} by the absorbance spectrum for the
large NQDs (provided by the NQDs distributer, N. N. Labs LLC.)
there is an energy level at $\sim$~570nm for the large NQDs with a
ground energy level at $\sim$~630nm. This was confirmed by PLE
experiments. The PLE graph, which is similar to the absorption graph,
shows strong absorption at $\sim570$nm with emission at
$\sim630$nm.

\subsection{\label{sec:level2}Resonant energy transfer between CdSe$/$ZnS NQDs as a function of separation}

Energy matched NQDs are used for the resonant energy transfer
experiments. The area of the PL signal of small NQDs is monitored
for any change. The small NQDs on the NSOM probe are optically
excited, and the induced excitons relax to their ground state
recombining and releasing a photon. These photons are collected
through the SIL and sent to a spectrometer generating the high energy
peak of the spectrum in Fig.~\ref{fig:PL+Abs+Levels}. This same
process also occurs with the large NQDs, generating the low energy peak
of the spectrum in Fig.~\ref{fig:PL+Abs+Levels}. As NQDs are
brought into close proximity, a portion of the small NQDs energy
would be expected to not be released through recombination and
transferred to the large NQDs. This interaction, associated to the
non-radiative energy transfer from the ground state of the small
NQDs to the third excited state of the large NQDs, becomes
increasingly more important as the separation between the NSOM
probe and the SIL decreases, \cite{Forster1948} within the
near-field region. Furthermore, since the intradot relaxation time
is very fast \cite{Efros1995,Sionnest1997,Klimov1999,Sionnest2005}
in the subpicosecond to picosecond range, energy transfer from the
large NQDs to the small ones is precluded.

The separation induced quenching of the small NQDs signal is a
clear signature of interaction between two groups of resonant
NQDs. Quenching of the small NQDs PL signal is shown in
Fig.~\ref{fig:PL}. The small NQDs PL signal decreases as a
function of separation: the area under the PL signal decreases
in the 15286, 15026, 14846, 14740, 14496, 14352 and 14006 sequence
fpr corresponding separations of 37, 35, 32, 30, 27,
23 and 18 nm respectively. Each PL spectrum in this experiment was
integrated over 120 seconds.

\begin{figure}
\begin{center}
\includegraphics[scale=0.33]{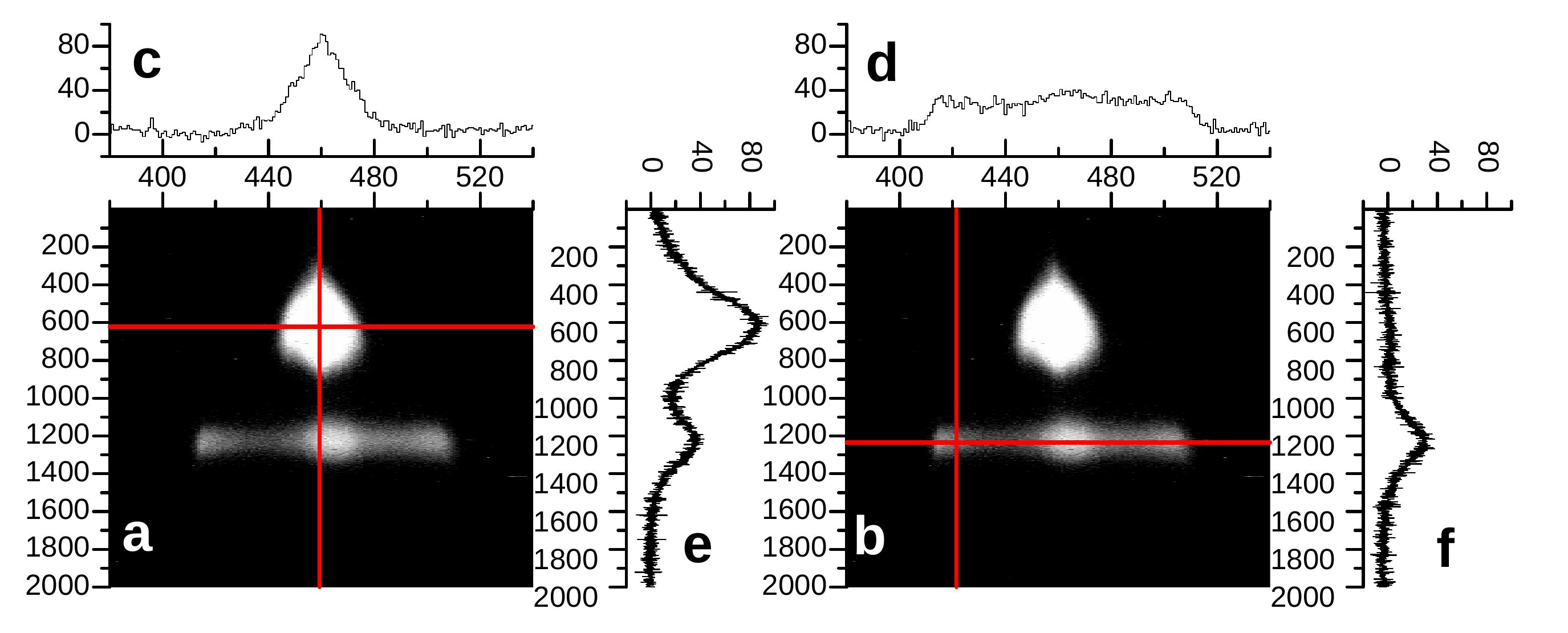}
\caption{(Color online) a, b) Images of small NQDs on the NSOM probe and large NQDs on the SIL are collected by the spectrometer.
These images show the distribution of photons as a function of energy on vertical axes. The horizontal axes show the spatial
distribution of NQDs. All axes are labeled by their pixel number.
c-f) Horizontal and vertical cross cuts from images a and b are shown on the top and right side of each image.
c) Horizontal cross cut of image a shows that small NQDs are clearly confined to the NSOM probe apex area (around pixel number 600).
d) Horizontal cross cut of image b shows that large NQDs are excited beyond the excitation area of NSOM probe on the SIL (around pixel number 1200).
e, f) Vertical cross cuts show the spectrum as function of energy.
\label{fig:Image+Both+QDs}}
\end{center}
\end{figure}

The corresponding enhancement in the large NQDs PL signal was not
observed as the number of NQDs on the SIL were not under control. As
a consequence of the deposition process of the large NQDs on the
SIL, many NQDs agglomerate. Lowering the concentration of large NQDs on
the 2.5 mm wide SIL did not prevent their agglomeration. This packed ensemble of monodispersed
large NQDs allows energy transfer between similar size neighboring \textbf{large}
NQDs, beyond the area above the NSOM probe.
Hence, numerous large \textbf{NQDs} get excited, as observed in the spectrometer images.
The spectrometer images in Fig.~\ref{fig:Image+Both+QDs} show that large NQDs are
excited beyond the area above the NSOM probe, strongly suggesting that neighboring
NQDs transfer energy to each other. Consequently, the area under the PL signal
of large NQDs is mostly constant.
\begin{figure}
\begin{center}
\includegraphics[scale=0.35]{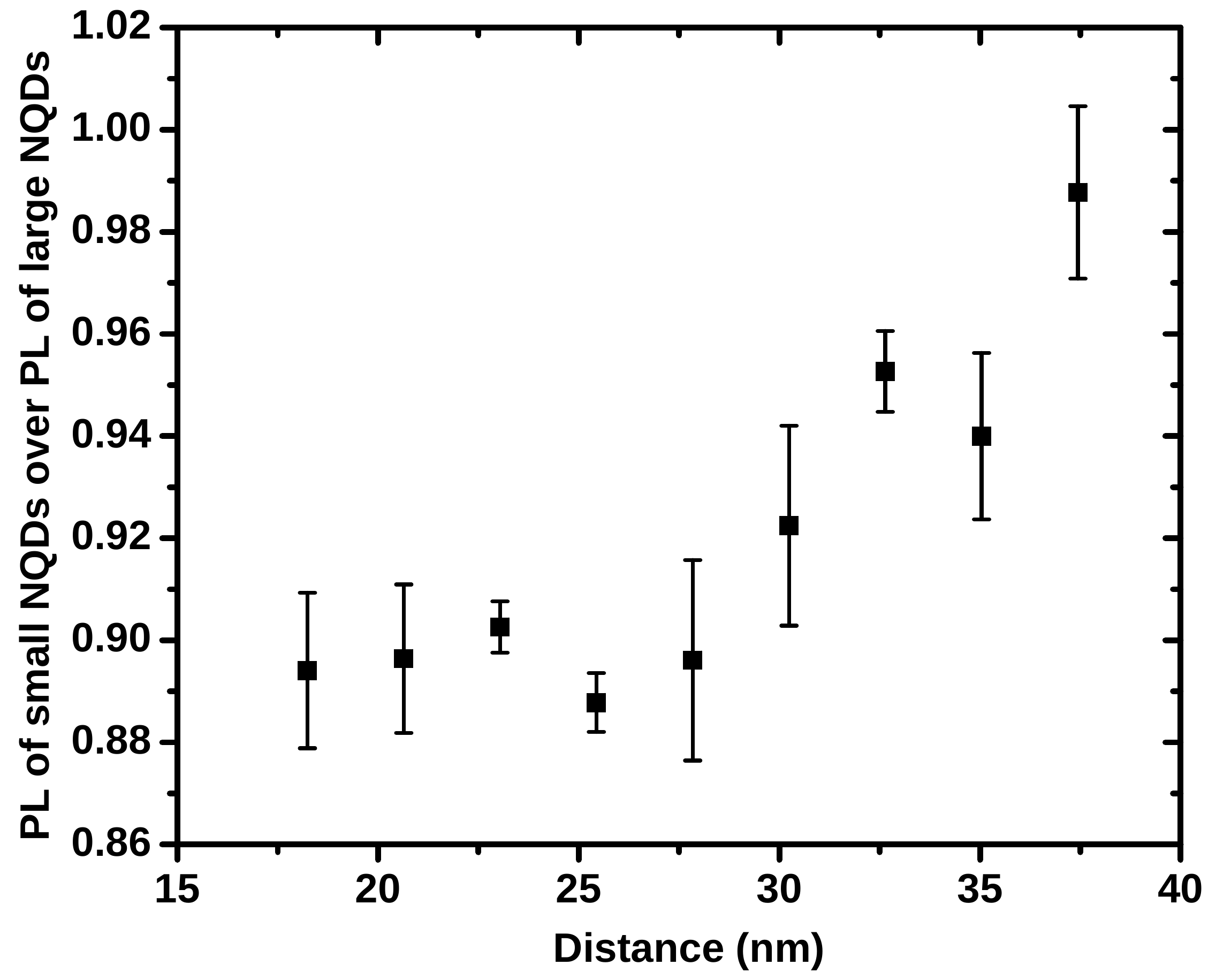}
\caption{Area under the small NQDs PL signal as a function of separation between
the NSOM probe and the SIL. These areas have been
normalized by the large NQDs signal areas. Experiment has been repeated 5 times for each point and
the standard errors have been used to calculate the error bars.\label{fig:Quench+PL+PRB}}
\end{center}
\end{figure}

Figure \ref{fig:Quench+PL+PRB}
shows the reduction of the PL signal of the small NQDs as a
function of separation between the NSOM probe and SIL. In this
figure the area under the PL signal of small NQDs has been
normalized to the area of the large NQDs PL signal. This last
normalization process is undertaken to cancel out small
fluctuations associated with the laser intensity. As it is shown
in the figure, when the separation reaches $\sim$20 nm, the
decrease in PL signal from the small NQDs stops, which is believed
to be the contact point between the two groups of NQDs on the NSOM
probe and the SIL. After this separation the PL signal of the
small NQDs remains constant. The Difference between this contact point and the sum of
the diameters of the two NQDs ($\sim$ 15 nm) could be partially due to the existence of
2.5 nm long and tightly bounded ODA ligands on the outer shell of NQDs which prevent full contact of NQDs.\cite{Belman2008,Li2003}
This separation is interpreted as a zeroth of the FRET spectroscopic ruler.

\begin{figure}[h]
\begin{center}
\includegraphics[scale=0.38]{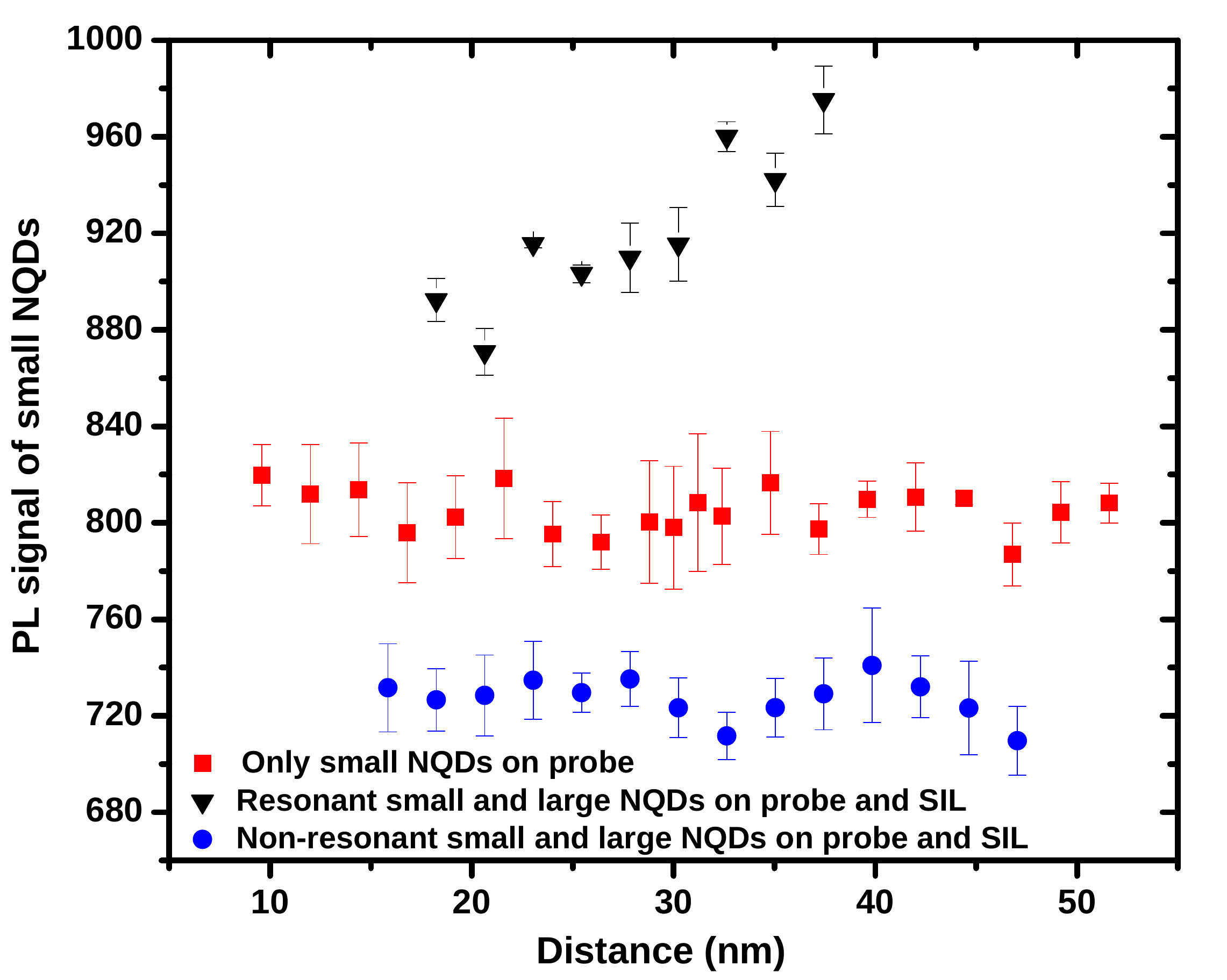}
\caption{(Color online) PL signal\textbf{s} of small NQDs while they are
interacting with resonant large NQDs (triangles, data shifted up for
clarity) show quenching as the NSOM probe approaches the SIL. PL signal\textbf{s} of small NQDs when there are no large NQDs
on the SIL (squares) do not show any changes as NSOM probe approaches the SIL. PL signal\textbf{s} of small NQDs in the
presence of non-resonance large NQDs do not show any changes (circles, data shifted for
clarity). All PL signal\textbf{s} have been normalized to laser intensity. Each experiment has been repeated 5 times for each point and
the standard errors have been used to calculate the error bars.\label{fig:quenching-nonquenching}}
\end{center}
\end{figure}

In addition the experiment was done without the large NQDs on the
SIL and only keeping the small NQDs on the NSOM probe. The PL signal of the
small NQDs does not change as the probe approaches the clean SIL,
as shown in Fig.~\ref{fig:quenching-nonquenching}. On the other hand, the triangle
points in Fig.~\ref{fig:quenching-nonquenching} show the quenching of the small NQDs PL signal
when both groups of small and large resonant NQDs were present.
To further enhance the argument that the change of PL signal in small NQDs is due
to FRET, nonresonant small and large NQDs were brought close
together. No quenching in the small NQDs PL signal was observed as it is shown by the circles in Fig. 6.
Small size NQDs were chosen to have difference ground energy level than any
levels of the large NQDs.

\section{\label{sec:level1}Theory}
\subsection{\label{sec:level2}Energy levels of the CdSe$/$ZnS}

Single band effective mass approximation
\cite{Kortan1990,Haus1993,Schooss1994} has been used to study the
excitonic energy levels of NQDs. This process helped us choose
the right resonant NQDs for the experiment. This numerical calculation
showed us that a large CdSe/ZnS with core radius of 3.7 nm and shell radius
of 4.25 nm with ground energy level of $3.135\times10^{-19}$ J (631 nm) has \textbf{its} third
excited energy level at $3.486\times10^{-19}$ J (567 nm). This make\textbf{s} it resonant
with a small CdSe/ZnS with outer radius of 3.25 nm with a ground energy level
emission at ~570 nm. In this calculation, the presence of ODA ligands on NQDs does not
change the energy levels.

The analysis is first restricted to the strong confinement
regime where the Coulomb interaction between particles is
neglected in comparison to the confinement energy. At the
core-shell boundary ($r=r_{c}$) the continuity of the wavefunction
and the probability current, \cite{Daniel1968,Brus1983} as well as
the boundary condition on the wavefunction at shell-vacuum
boundary of the NQDs (assumed to be immersed in vacuum) yield the
energy levels of the free particles (electrons and holes).
The shell radius of the NQDs $r_{s}$ is obtained from small
angle X-ray diffraction scattering. Knowing $r_{s}$, the core radius $r_{c}$ can
be adjusted to change the ground energy level of the calculation, which
is then used to find other energy levels.

Later the electron-hole Coulomb interaction energy is considered a
correction to the total Hamiltonian. This last term is small
and is treated as a Helium-like perturbation\cite{Schooss1994} for
the electron and hole energy of the system. At this point, by adjusting
$r_{c}$, the ground state energy due to strong confinement
and electron hole Coulomb interaction correction can be calculated and
compared to the observed PL peak of the NQDs. Matching the calculated energy
of the ground state and the observed PL peak leads us to choose the right
$r_{c}$. For $r_{c} = 3.7$ nm and $r_{s} = 4.25$ nm the first
four energy levels and their Coulomb correction are shown in Table {\ref{tab:energy}}.
The calculated excited energy levels are compared to the absorption peaks of
the dots, showing very good agreement.

\begin{table*}
\caption{Table of calculated confinement energy levels and Coulomb
correction terms for CdSe/ZnS NQDs with $r_{c}$=3.7 nm and
$r_{s}$=4.25 nm. $\ell_{e}$ and $\ell_{h}$ are quantum numbers
of electrons and holes inside the NQDs.
The last column is the calculated
wavelength for the excitonic recombination. In addition to
effective masses and band gap, the conduction bands offset between
CdSe and ZnS were used for these calculations.
\cite{Miklosz1967,Dimmock1962,Nethercot1972}}

\begin{tabular}{l l l l l l l l@{ } }
 $\ell_{e}$&$\ell_{h}$&$~~E_{e}(J)$&$~~E_{h}(J)$&$~~E_{c}(J)$&$~~E_{Total}(J)$&{ }$~~\lambda(nm)$\\
 \hline
 0&0&~~$1.699\times10^{-19}$&~~$1.554\times10^{-19}$&~~$-1.22\times10^{-20}$&$~~3.135\times10^{-19}$&$~~631$\\
 0&1&$~~1.699\times10^{-19}$&~~$1.643\times10^{-19}$&~~$-1.13\times10^{-20}$&$~~3.233\times10^{-19}$&$~~612$\\
 1&0&$~~1.943\times10^{-19}$&~~$1.554\times10^{-19}$&~~$-1.17\times10^{-20}$&$~~3.384\times10^{-19}$&$~~584$\\
 1&1&$~~1.943\times10^{-19}$&~~$1.643\times10^{-19}$&~~$-9.60\times10^{-21}$&$~~3.486\times10^{-19}$&$~~567$\\
\end{tabular}
\label{tab:energy}
\end{table*}

\subsection{\label{sec:level2}Dipole-dipole interaction and resonant energy transfer}

The energy of any charge distribution in the presence of other
charge distributions and external electrical potential can be
obtained by a multipolar expansion.\cite{Jackson} Since both
NQDs are neutral, the first term which survives is the
dipole-dipole interaction energy, due to the dipolar electric
field of one of the excitons, interacting with the other NQDs
exciton's dipole.

This energy of interaction between electric multipoles may be
found by expanding the Coulomb interaction. First consider two
charge distributions, 1 and 2, centered at $O_{1}$ and $O_{2}$ respectively with
coordinate axes chosen to be parallel. The distance between these
two origins is defined as $R$ which makes an angle $\theta$ with
the $z$ axis of the first charge distribution. The separation
between two elements, $i$ and $j$, of these two charge distributions
is defined as $r_{ij}$. By expanding this distance into spherical
harmonics, the electrostatic interaction can be written as
\cite{Carlson1950,Wolf1968}

\begin{eqnarray}
V_{12}=\sum_{i,j}(\frac{e^{2}}{r_{ij}}) = \frac{1}{4 \pi \epsilon} e^{2}\sum_{i,j}\sum_{\ell,\ell^{'}}\frac{(-1)^{\ell^{'}}r_{i}^{\ell}r_{j}^{\ell^{'}}}{R^{\ell+\ell^{'}+1}} {}\nonumber\\
\times \sum_{m,m^{'}}B_{\ell\ell^{'}}^{m
m^{'}}Y_{\ell+\ell^{'}}^{-m-m^{'}}(\theta,0)Y_{\ell}^{m}(\theta_{i},\phi_{i})Y_{\ell^{'}}^{m^{'}}(\theta_{j},\phi_{j}),
\label{eq:V}
\end{eqnarray}
\noindent where
\begin{eqnarray}
B_{\ell\ell^{'}}^{m m^{'}}= \frac{(-1)^{m+m^{'}}(4\pi)^{\frac{3}{2}}}{[(2\ell+1)(2\ell^{'}+1)(2\ell+2\ell^{'}+1)]^{\frac{1}{2}}} {}\nonumber\\
\times (\frac{(\ell+\ell^{'}+m+m^{'})!(\ell+\ell^{'}-m-m^{'})!}{(\ell+m)!(\ell-m)!(\ell^{'}+m^{'})!(\ell^{'}-m^{'})!})^{\frac{1}{2}}.
\label{eq:B}
\end{eqnarray}

For the case of interaction between two NQDs, $i$ is the charge
distribution of the first NQD and $j$ belongs to the second NQD.
When the ground state of the small NQD ($\ell_{e}=0,\ell_{h}=0$)
is in resonance with the third excited state of the large NQD
($\ell_{e}=1,\ell_{h}=1$), the emission peak of the small NQD
overlaps with the fourth absorption peak of the large NQD. Both of
these states are optically active due to the P symmetry in the
valence band.

\begin{figure}
\begin{center}
\includegraphics[scale=0.33]{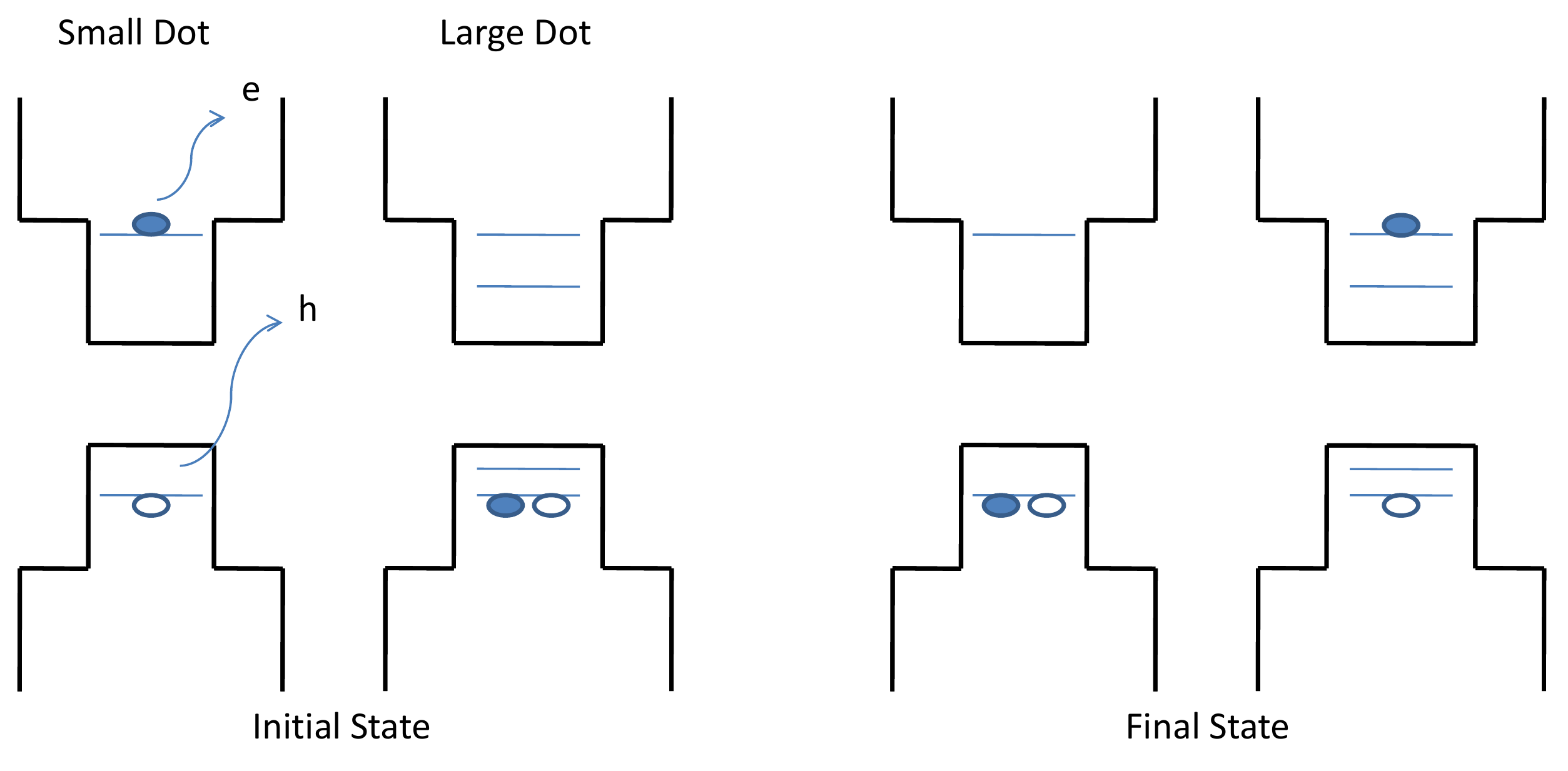}
\caption{Initial and final states of combination of small and large NQDs before and after the energy transfer. In the initial state, the small NQD exciton is in its ground state and no exciton exists in the large NQD. In the final state, there is no exciton in the small NQD and the large NQD exciton is in its third excited state.\label{fig:interaction}}
\end{center}
\end{figure}

We have used Eqs. (\ref{eq:V}) and (\ref{eq:B}) to calculate the
transition rate $W=\frac{2\pi}{\hbar} |<V_{12}>|^{2} \rho$. \cite{Scholes2003,Curutchet2008,Baer2008,Scholes2005,Allan2007}
$<V_{12}>$ is the Coulomb potential energy between the small and
large NQDs and $\rho$ is the normalized overlap between donor
emission and acceptor absorption
spectra.\cite{Scholes2003,Curutchet2008,Baer2008,Scholes2005} From
this equation,  the transition probabilities $P=W\tau$ are
obtained, with $\tau$ the donor's lifetime. Since the exciton at
the ground state on the small NQDs recombines after $\tau$, the
energy transfer between two resonant NQDs happens in times shorter
than $\tau$.
The overall initial wavefunction is the multiplication of the
wavefunction of the exciton (electron-hole pair) at its ground
state in the small NQD by the wavefunction
of no exciton in the large NQD (which is
equivalent to have an electron and hole both in the first
excited energy level in the valence band of the large NQD).
Similarly the overall final wavefunction is considered to be the
product of the wavefunction of no exciton in the small NQD by the wavefunction
of an exciton at the third excited energy state in the large NQD.
Fig.~\ref{fig:interaction} shows the initial and final states
considered.
Using these initial and final states, $<V_{12}>=\frac{1.19\times 10^{-46}}{r^{3}}
J$ is obtained. Here the normalized overlap between donor's
emission and acceptor's absorption, $\rho=3.09\times10^{19}$,
obtained from the experimental data, has been used. Hence, the
transition rate is: $W=\frac{2\pi}{\hbar} <V_{12}>^{2}
\rho=(2.60\times10^{-38}\frac{m^{6}}{s})/r^{6}$ and the
transition probability is
$P=W\tau=(2.60\times10^{-47}m^{6})/r^{6}$ by considering $\tau =
1$ nanosecond. Radiative lifetime of CdSe or CdSe/ZnS has been
measured or calculated to be in the range of few nanoseconds up to
several tens of nanoseconds.
\cite{Efros1992,Michler2000,Dahan2001,Zhang2002,Schlegel2002,Wang2003,Klimov1999,Matsumoto2007,Fisher2004}

\section{\label{sec:level1}Discussion}

\begin{figure}[t]
\begin{center}
\includegraphics[scale=0.35]{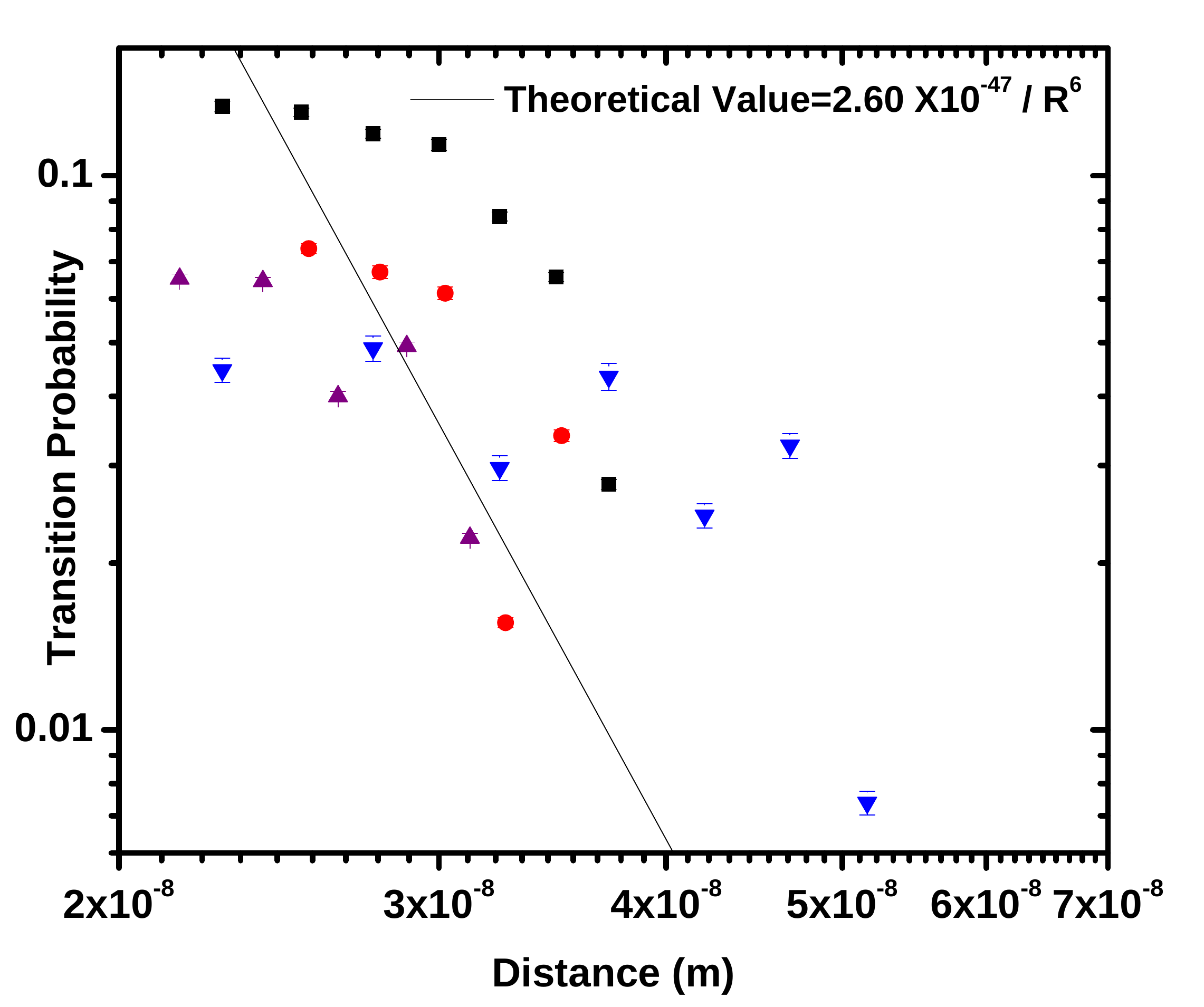}
\caption{(Color online) Transition probability from 4 different experiments and
theoretical calculation as a function of separation. The vertical
axis represents the transition probability, obtained by
subtracting the normalized small NQDs PL signal at the farthest
experimental point from the normalized small NQDs PL at measurement
point, divided by the normalized PL signal at the farthest
distance. The solid line shows the theoretical value for the
transition probability calculated in section III-B using a dipole-dipole approximation.\label{fig:PT}}
\end{center}
\end{figure}

Data similar to those reported in Fig. \ref{fig:PL}
have been used to derive the
transition probability of the resonant energy transfer between
these two groups of NQDs. The procedure to obtain the transition
probability is given by

\begin{equation}
P=\frac{A(\infty) - A(r)}{A(\infty)}, \label{eq:Prob}
\end{equation}

\noindent where $A(r)$ is the normalized area of the PL signal of
the small NQDs at a separation $r$ and $A(\infty)$ is the
corresponding one at an infinite separation, when there is no
interaction between the small and large dots.
Equation (\ref{eq:Prob}) represents the fact that as the two groups
of NQDs get closer the dipole-dipole interaction
increases and the probability of resonant energy transfer
increases. Hence, the normalized PL signal of the small NQDs
decreases proportionally to the square of the strength of the
interaction. In the experimental case $A(\infty)$ has been
selected at the distance when the interaction is the smallest,
i.e. the largest experimentally accessible separation between
groups of NQDs. Since the PL signal of small NQDs cannot be
collected when they are very far away from the SIL, a position
where the PL signal from small NQDs is completely collected has to
be used as a reference. Furthermore, the feedback interaction
between the NSOM probe and the SIL happens over $\sim$~50~nm,
which limits the point for the largest separation to about this
value. As an example, for the experiment that
Fig. \ref{fig:Quench+PL+PRB} was extracted, this distance is 37
nm.

Figure \ref{fig:PT} shows the transition probability for four
experimental sets. The theoretical value for the transition
probability ($2.60 \times10^{-47} m^{6} /r^{6}$) is also shown in
the figure.

The transition probability also provides the F\"{o}rster radius as
it is represented by \textrm{$R_{o}$} in the F\"{o}rster rate
equation.\cite{Forster1948,Forster1965,JaresErijman2003,Muller2004}
By comparing \textrm{$(R_{o}/r)^{6}$} from the F\"{o}rster rate equation
and its equivalent experimental transition probability, the
F\"{o}rster radius is calculated to be in the 14-22 nm range.
From the theoretically calculated transition probability, \textbf{$R_{o}$} radius is extracted to be 17 nm.
A F\"{o}rster radius of 4.7 nm was obtained by C. Kagan \textit{et al.}\cite{Kagan1995,Kagan1996}
using differently sized CdSe NQDs and capping ligands, under a closed packed mixture of two sizes of
NQDs. In our experiment the measurement is between small and large NQDs that are isolated from each
other, while in Ref. [13, 14] it is between mixed small and large NQDs.
The authors of Ref. [13, 14] have used spectral overlap of donor emission and acceptor absorption
integral to measure F\"{o}rster radius. C. Kagan et al. show \cite{Kagan1995,Kagan1996,Dexter1953}

\begin{equation}
R_{o}^{6}\propto  \frac{\varphi_{D}}{n^{4}} \int^{\infty}_{0} F_{D}(\nu) \varepsilon_{A}(\nu) \frac{d\nu}{\nu^{4}}
\end{equation}

\noindent $\nu$ is the frequency, $\varphi_{D}$ is the donor luminescence
quantum yield, and $n$ is the effective index of refraction. $F_{D}(\nu)$ is the normalized
spectrum for the donor and $\varepsilon_{A}(\nu)$ is the molar extinction coefficient for
acceptor\textbf{s} absorption. The authors used a random closed packed mixture of NQDs with organic
caps filling interstices and considered the volume weighted average of the index of refraction of CdSe ($n=2.58$)
and organic caps ($n=1.47$) as an effective index of refraction. In our experiment, isolating small NQDs
from large NQDs would make $n$ smaller in comparison to these works because of the presence of air between the interacting NQDs.
This screening effect has been discussed previously in similar systems. \cite{Beljonne2009}
Using $n$, spectral overlap and $\varphi_{D}$ under our experimental condition
would bring the F\"{o}rster radius obtained from Ref. [13, 14] in close agreement with our data.
By considering the parameters for L. Guo \textit{et al}. work, \cite{Guo2006} a
similar conclusion can be obtained.

\section{\label{sec:level1}Conclusion}

In conclusion, resonant energy transfer between two groups of
CdSe/ZnS as a function of separation has been observed directly
from the PL signal of small NQDs.
Small CdSe/ZnS NQDs on the apex of the NSOM probe were
brought into close proximity to the resonant large NQDs on the SIL
and both groups of NQDs were optically excited. As the third
excited state energy level of large NQDs is the closest energy
level to the ground state energy level of the small NQDs, some
fraction of the energy was transferred from small NQDs to the large
NQDs before recombination took place in the small NQDs.
Figure \ref{fig:PT} indicates that the interaction between resonant
NQDs could be a dipole-dipole interaction.
Within the experimental resolution, this energy transfer is compatible
with a dipole active one and depends on distance as dipole-dipole
interaction ($\propto \frac{1}{r^{6}}$). In the future, more work
will be needed to isolate a single small NQD on the NSOM probe and a
single large NQD on the SIL.

In all experiments, the small PL signal reaches a point that the
quenching stops and the PL signal becomes constant, which seems to
correspond to the contact point of the small and large NQDs as
they get close enough to each other. For all of the experiments
this contact point is $\sim$~20 nm, comparable to the sum of the
size of the two NQDs diameters of $\sim$~15 nm. This difference is
mostly due to the presence of the 2.5 nm long, tightly bounded ODA ligands on the NQDs.

The experimental transition probability between $\sim $6.5 nm and
$\sim $8.5 nm diameter CdSe/ZnS NQDs is measured in the range of
$(0.7\times10^{-47}m^{6})/r^{6}-(11.0\times10^{-47}m^{6})/r^{6}$,
while the theoretically calculated value is
\textrm{($2.60\times10^{-47}m^{6} /r^{6}$)}. Figure \ref{fig:PT}
indicates that the interaction between resonant NQDs could be a
dipole-dipole interaction. More precise data is needed for obtaining a better
fitting. Comparison of the F\"{o}rster radius from our experiment, 14-22 nm,
with the distance between NSOM probe and SIL, (15-50 nm), shows that coupling
between NQDs is a near-neighbor interaction.

\begin{acknowledgments}
This work was supported in part by the National Science Foundation through grants No.
CCF-0508239 and PHY-0701636, and Los Alamos National Laboratory support through contract No.
49423-001-07. The authors are also indebted to the Nanoscale
Imaging Center at IUPUI for the liberal use of the installations.
We are also indebted to Dr. Horia Petrache for conducting the
NQDs size measurement by small angle X-ray scattering and Cynthia
Wassall for preparing the SIL.
\end{acknowledgments}

\bibliography{unsrt}

\end{document}